%% file: reduced.tex
\documentclass[sigconf,nonacm,natbib=false]{acmart}

\AtBeginDocument{%
  }

\setcopyright{acmlicensed}
\copyrightyear{2018}
\acmYear{2018}
\acmDOI{XXXXXXX.XXXXXXX}

\acmConference[Conference acronym 'XX]{Make sure to enter the correct
  conference title from your rights confirmation emai}{June 03--05,
  2018}{Woodstock, NY}
\acmISBN{978-1-4503-XXXX-X/18/06}

\RequirePackage[
  datamodel=acmdatamodel,
  style=acmnumeric,
  ]{biblatex}

\usepackage{amsmath}
\usepackage{algorithm}
\usepackage{algpseudocode}
\usepackage{hyperref}
\usepackage{listings}
\usepackage{xcolor}

\usepackage{graphicx} 
\usepackage{listings}
\usepackage{xcolor}
\usepackage{fancyvrb}

\date{December 10, 2023} 

\begin{document}
\title[Augmenting speech transcripts of VR recordings]{Augmenting speech transcripts of VR recordings with gaze, pointing, and visual context for multimodal coreference resolution}

\author{Riccardo Bovo}
\affiliation{%
\institution{Imperial College London}
  \city{London}
  \country{United Kingdom}}
\email{rb1619@ic.ac.uk}
\orcid{1234-5678-9012}

\author{Frederik Brudy}
\email{frederik.brudy@autodesk.com}
\orcid{0000-0002-3868-0967}

\affiliation{
 \institution{Autodesk Research}
  \city{Toronto}
  \state{Ontario}
  \country{Canada}
}

\author{George Fitzmaurice}
\email{george.fitzmaurice@autodesk.com}
\orcid{0000-0002-2834-7757}

\affiliation{
 \institution{Autodesk Research}
  \city{Toronto}
  \state{Ontario}
  \country{Canada}
}

\author{Fraser Anderson}
\email{fraser.anderson@autodesk.com}
\orcid{0000-0003-3486-8943}

\affiliation{
\institution{Autodesk Research}
  \city{Toronto}
  \state{Ontario}
  \country{Canada}
}

\renewcommand{\shortauthors}{Bovo et al.}

\definecolor{changeNoteColorCHI}{rgb}{0.1,0.6,1}
\newcommand{\revisedCHI}[1]{\textsf{\textbf{\textcolor{changeNoteColorCHI}{#1}}}} 

\definecolor{changeNoteColorUIST}{rgb}{0.9,0.3,1}
\newcommand{\revisedUIST}[1]{\textsf{\textbf{\textcolor{changeNoteColorUIST}{#1}}}}  

\definecolor{changeNoteColor}{rgb}{1,0.6,0}
\newcommand{\revised}[1]{\textsf{\textbf{\textcolor{changeNoteColor}{#1}}}}  

\newcommand{\generalRE}[1]{\colorbox{black!20!white}{#1}}
\newcommand{\spatialRE}[1]{\colorbox{green!20!white}{#1}}
\newcommand{\spatialimplicitRE}[1]{\colorbox{purple!20!white}{#1}}
\newcommand{\spatialexplicitRE}[1]{\colorbox{orange!20!white}{#1}}
\newcommand{\exophora}[1]{\colorbox{yellow!20!white}{#1}}
\newcommand{\REendophora}[1]{\colorbox{blue!20!white}{#1}}
\definecolor{orangishred}{RGB}{255, 100, 50} 
\definecolor{greenbluecyan}{RGB}{0, 200, 170} 
\newcommand{\correct}[1]{\colorbox{greenbluecyan!20!white}{#1}}
\newcommand{\wrong}[1]{\colorbox{orangishred!20!white}{#1}}
\newcommand{\argumented}[1]{\colorbox{blue!10!white}{#1}}
\newcommand{\correctt}[1]{\colorbox{green!20!white}{#1}}
\newcommand{\wrongg}[1]{\colorbox{red!20!white}{#1}}

\begin{abstract}
Understanding transcripts of immersive multimodal conversations is challenging because speakers frequently rely on visual context and non-verbal cues, such as gestures and visual attention, which are not captured in speech alone. 
This lack of information makes coreferences resolution-the task of linking ambiguous expressions like ``it'' or ``there'' to their intended referents-particularly challenging.
In this paper we present a system that augments VR speech transcript with eye-tracking laser pointing data, and scene metadata to generate textual descriptions of non-verbal communication and the corresponding objects of interest. 
To evaluate the system, we collected gaze, gesture, and voice data from 12 participants (6 pairs) engaged in an open-ended design critique of a 3D model of an apartment. 
Our results show a 26.5\% improvement in coreference resolution accuracy by a GPT model when using our multimodal transcript compared to a speech-only baseline. 
\end{abstract}

\ccsdesc[500]{Human-centered computing}
\ccsdesc[500]{Human-centered computing~Natural language interfaces}
\ccsdesc[500]{Human-centered computing~Virtual reality}
\ccsdesc[500]{Human-centered computing~Collaborative interaction}

\keywords{multimodal coreference resolution, virtual reality (VR), speech, gaze, pointing}

\begin{teaserfigure}
  \includegraphics[width=\textwidth]{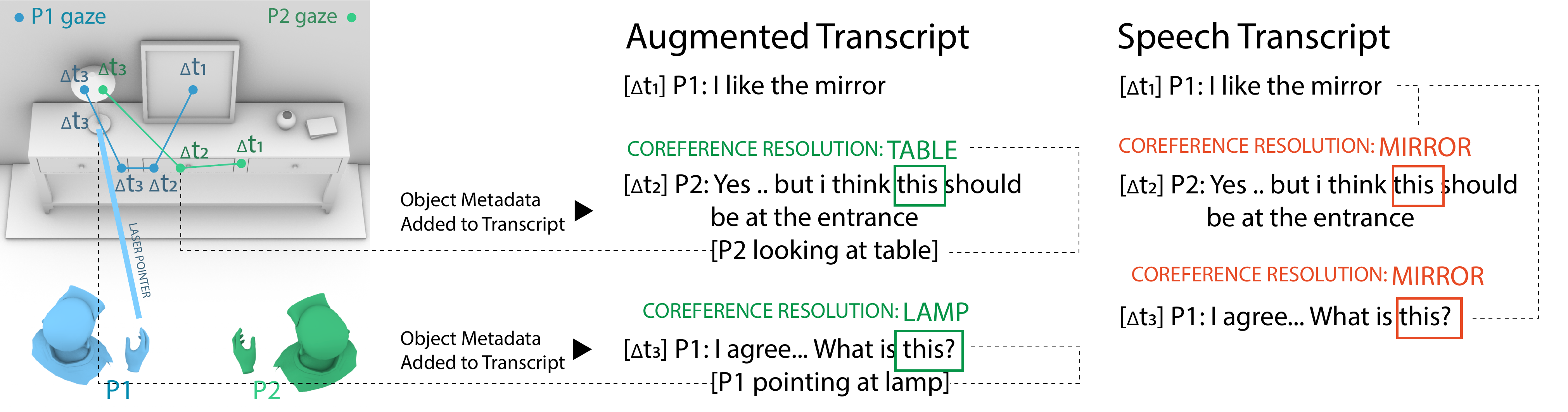}
  \caption{Depiction of our system performing coreference resolution by leveraging non verbal cues such as pointing and visual attention. The system uses user's pointing behaviour and eye-gaze to determine the referent of ambiguous referring expressions, by correlating the location of eye gaze and pointing targets with pronoun "this" used by users. }
  \label{fig:teaser}
\end{teaserfigure}

\received{20 February 2007}
\received[revised]{12 March 2009}
\received[accepted]{5 June 2009}

\maketitle

\input{1-introduction-reduced}

\input{2-related_work-reduced}


\input{3-system_design_reduced}

\input{4-data_collection_reduced}

\input{5-result-reduced}

\input{6-discussion-reduced}

\printbibliography


\end{document}

%% file: 1-introduction-reduced.tex
\section{Introduction}

The use of VR and AR for collaborative applications is rapidly growing, enabling immersive recordings of activities, such as design reviews \cite{liu2020evaluating,liu2014virtual,lopez2018virtual,thewild,workshopxr,arkio}, remote support \cite{realwear,Kim2020}, and social interactions \cite{vrchat}. These recordings are increasingly revisited to support collaboration, analysis, and follow-up tasks. As adoption grows, machines will need to better understand such recordings in order to summarize discussions, extract insights, and provide accessibility support. 



However, immersive conversations (both, in real-world or mixed reality environments) are inherently multi-modal. Besides speech, conversation participants frequently use non-verbal cues--such as gaze and pointing--to establish shared understanding (\autoref{fig:teaser}). This poses a difficulty for machine comprehension. \textit{Referring expressions} (REs), such as ``this'' or ``that,'' are common in natural conversation, yet their meaning often depends on visual or gestural context. For example, in the utterance ``I like it!'' the RE ``it'' can only be understood when paired with a gesture or gaze toward the referent.
Current Large Language Models (LLMs) and Visual Large Language Models (VLMs) can process transcripts of meetings to generate summaries and insights, but they struggle with these ambiguous Referring Expressions, leading to incorrect or incomplete identification of referenced objects. This limits the ability of intelligent systems to support tasks such as summarizing VR conversations \cite{Zhang2023ConceptEVA:Summaries}, extracting information from design review meetings \cite{Mahadevan2023Tesseract:Miniature}, and offering real-time accessibility support. 



Prior research has explored grounding conversation in shared visual contexts by linking dialogue to entities in a scene using neural models \cite{Yu2019WhatDialogues, Goel2022WhoNarrations, KongWhat2014WhatAre, GuoGRAVL-BERT:Resolution, Kottur2021SIMMCConversations}. However, these approaches are ineffective when speakers never explicitly name the object they are referring to (exophora). For example, in the dialogue `P1: "What do you think about this?"` `P2: "It does not look comfortable!"`, the referent remains unclear without additional cues. Even when entities are explicitly named (endophora), coreference resolution is often unreliable in natural conversations \cite{Zhang2023ConceptEVA:Summaries}. For instance, after `P1: "There is a funny coffee table."` a response like `P2: "We should move this."` could refer either to the table or another object introduced later. While seminal HCI work such as ``Put That There'' \cite{10.1145/800250.807503} and follow-ups \cite{romaniak2020nimble, Miniotas2006SpeechGaze, Bovo2023Speech-AugmentedAnalysis} leveraged non-verbal cues in human–machine interaction, they did not address multi-speaker conversations or use such cues for transcript comprehension. Similarly, studies on gaze synchrony in collaboration \cite{Vrzakova2019, Villamor2018, Moulder2023, Pietinen2008} highlight its role in establishing shared focus, but have not quantified how effectively it identifies the precise object under discussion.


This paper proposes a system (\autoref{fig:system}) that augments VR speech transcripts to improve coreference resolution, by leveraging non-verbal cues. By inferring speakers' attention through their gaze and pointing behavior, the system disambiguates spatial implicit referring expressions (REs) such as `P1: "What do you think about this?"` or `P2: "It does not look comfortable."' The system augments the transcript with contextual cues, linking these expressions to the intended object in the scene (e.g., the living room sofa), enabling more accurate downstream processing, such as summarization and information extraction. Our primary contribution is a system that integrates gaze and pointing into VR speech transcripts to address the fundamental problem of coreference resolution. We built a multi-user VR application that captures verbal interactions, eye-tracking data, and laser-pointing behavior. VR provides a controlled environment where pre-segmented 3D objects simplify linking verbal references and non-verbal behavior, avoiding the challenges of object detection and segmentation that arise in AR or real-world settings. VR is increasingly adopted for collaborative design review in both academia \cite{liu2020evaluating, liu2014virtual, lopez2018virtual} and commercial applications \cite{thewild, workshopxr, arkio}, making it a natural setting to systematically study how non-verbal cues can be integrated into transcript comprehension. To evaluate our system, we conducted a study with 12 participants (six pairs) performing a collaborative design review in VR. We compared coreference resolution performance performed by Chat GPT4 on our augmented transcripts against a baseline using only speech transcripts.



The primary contributions of this work are twofold:
\begin{itemize}
    \item A novel \textbf{system} that augments speech transcripts from collaborative VR sessions with non-verbal cues (gaze and pointing) to resolve ambiguous referring expressions, validated through a 12-participant study that demonstrates a \textbf{26.5\% improvement} in coreference resolution accuracy over a speech-only baseline.
    \item A \textbf{quantitative analysis} of how different non-verbal behaviors---gaze and pointing---and their synergies (individual, concurrent, and recurrent patterns) contribute to identifying the object of interest, establishing a clear hierarchy of cues for resolving ambiguity in immersive collaborative dialogue.
\end{itemize}

%% file: 2-related_work-reduced.tex
\section{Related Work}
We review previous work on \textit{coreference resolution}, \textit{visual attention} as well as \textit{pointing based communication} in collaborative settings, and how HCI research leveraged comprehending deictic behaviour such as speech+gaze and speech+gestures through \textit{non verbal and multimodal interaction}.

\subsection{Coreference Resolution}

\textit{Coreference resolution} involves identifying words and phrases in a text that refer to the same entity, a crucial task in natural language processing \cite{Goel2022WhoNarrations}. This task is particularly challenging in conversational contexts due to their fluid structure, dynamic topic shifts, ambiguous pronoun use, and implicit shared knowledge, which can obscure clear reference resolution \cite{Yu2019WhatDialogues}. For example, speakers often use pronouns instead of specific names, complicating entity linking \cite{bai2021}.

\subsubsection{Visual Coreference Resolution}

Recently, coreference resolution has significantly advanced through machine learning, especially when integrating visual and textual data to link text expressions to entities in images. This involves a system's ability to understand linguistic cues and visual features, using Neural Networks to process visual scenes and identify entities connected to text mentions \cite{Goel2022WhoNarrations, Yu2022VD-PCR:Resolution, Yu2019WhatDialogues}. \citeauthor{Yu2019WhatDialogues} developed \textit{VisCoref}, a model for visual pronoun coreference resolution using deep learning techniques \cite{Yu2019WhatDialogues}. \citeauthor{Goel2022WhoNarrations} utilized "weak supervision" to train a model that identifies coreferences in text-image pairs and determines pronoun referents \cite{Goel2022WhoNarrations}. \citeauthor{Yu2022VD-PCR:Resolution} introduced \textit{VD-PCR}, a framework to improve Visual Dialog comprehension through Pronoun Coreference Resolution by training a multi-modal BERT to understand pronouns in image-dialogue pairs and pruning dialogue to retain relevant input \cite{Yu2022VD-PCR:Resolution}. These approaches involve resolving coreference by training neural networks to analyze visual scenes, using one or multiple neural networks to process 2D images for visual scene analysis.

\subsubsection{Immersive Visual Coreference Resolution}


Several studies have approached visual coreference resolution using 3D visual representations instead of 2D. 
\citeauthor{KongWhat2014WhatAre} \cite{KongWhat2014WhatAre} presented a method that uses natural language descriptions of RGB-D scenes to enhance visual scene comprehension. \citeauthor{Hong20233D-LLM:Models} advanced this by embedding 3D world knowledge into expansive language models in their \textit{3D-LLM} \cite{Hong20233D-LLM:Models}, providing insights into coreference resolution in three-dimensional contexts. These methods resolve coreferences by analyzing the visual scene, disambiguate entities, and correlating them with textual mentions.
\citeauthor{Kottur2021SIMMCConversations} introduced the SIMMC 2.0 dataset, which includes immersive multi-modal conversations and a baseline model for coreference resolution and multi-modal disambiguation to improve AI assistants \cite{Kottur2021SIMMCConversations}.  Dynamics of coreference resolution in human-AI interactions inherently differ from those in human-human exchanges, where in human-AI dialogues, the AI system can request clarifications when visual coreference resolution is uncertain, while seeking clarification is inherently limited when analysing human-human dialogues, especially when the original speakers are no longer accessible for further context. Using the SIMMC 2.0 dataset, \citeauthor{GuoGRAVL-BERT:Resolution} proposed a framework that uses metadata in the field of view to help disambiguate objects and correlate them with textual dialogue information \cite{GuoGRAVL-BERT:Resolution}.


We use a similar approach by extracting metadata from the scene based on a user's nonverbal cues (e.g., gaze, pointing). Previous methods resolve coreferences by analyzing the visual scene with neural networks, which involves understanding the scene, segmenting entities, and modeling relationships with mentions in the accompanying text. In contrast we adopt a simpler approach to visual coreference resolution by leveraging non-verbal cues (e.g., users' visual attention and pointing) to identify the target in a 3D scene, enhancing a speech transcript with contextual information.

\subsection{Visual attention during communication}
\label{visattent}

Synergies of gaze are common during human-human collaboration, stemming from the shared visual context but extending beyond visual alignment. Previous research highlights how this fosters mutual understanding, reduces the likelihood of misunderstandings, and enhances collaboration. \citeauthor{Vrzakova2019} used recurrence quantification analyses (RQA) to identify patterns in visual attention during collaborative tasks, showing alignment with screen activity correlated with team performance and collaboration \cite{Vrzakova2019}. \citeauthor{Villamor2018} found concurrent visual attention crucial in pair-programming tasks \cite{Villamor2018}, while \citeauthor{Moulder2023} quantified team-level gaze dynamics using RQA, finding them predictive of task success \cite{Moulder2023}. Awareness of visual attention helps ground referring expressions (REs) within a visual scene. \citeauthor{Schneider2013} demonstrated enhanced collaboration with mutual gaze perception in 2D tasks \cite{Schneider2013}, and \citeauthor{Zhang2017} showed gaze cursors facilitated communication in 2D screen interactions \cite{Zhang2017}. \citeauthor{DAngelo2017} found visual attention cues reduced communication complexity in pair programming, aiding implicit referring expressions \cite{DAngelo2017}. Similarly, visual attention can help to ground context in immersive scenarios where the visual context spans 360 degrees, creating blind spots for interlocutors known as the fragmentation problem. \citeauthor{Hindmarsh1998FragmentedEnvironments} introduced this in collaborative environments \cite{Hindmarsh1998FragmentedEnvironments}, and \citeauthor{Bovo2022} showed bidirectional head-based cues in VR increased mutual awareness and reduced cognitive load during data analysis \cite{Bovo2022}. \citeauthor{Jing2021} found bidirectional eye gaze cues improved co-presence, gaze awareness, and collaboration in MR environments \cite{Jing2021}.

This prior work has shown the importance of visual attention awareness in simplifying human-human dialogue comprehension by enabling more implicit communication. In our work, we leverage visual attention to enhance machine comprehension of immersive human-human dialogues. Visual attention patterns, especially in tasks where participants work closely together, have been shown to predict various aspects of the collaboration, ranging from task success to the quality of the collaborative experience. However, to the best of our knowledge, visual attention synergism, such as concurrent and recurrence of visual attention, has never been used as a retrieval method to identify the object of attention during collaborative discussion.

\subsection{Pointing Based Communication}
\label{pointingbasedcom} 


The absence of visual attention awareness, the complexity of a visual scene, and the difficulty of verbally describing a referent prompt the use of pointing gestures with utterances, known as deictic expressions \cite{Wong2010}. Research shows that pointing gestures, particularly in immersive multi-modal conversations, significantly aid in referent disambiguation \cite{Wong2010}. In AR/VR, pointing gestures become even more effective due to the use of lasers, enabling users to specify the referent of a pointing gesture \cite{Wong2014SupportDeicticPointing}.


\citeauthor{Piumsomboon2019} found that virtual awareness cues, including pointing gestures, field of view, and eye gaze, significantly improved user performance, usability, and subjective preferences in MR collaboration \cite{Piumsomboon2019}. Similarly, \citeauthor{Bovo2022} demonstrated that users prefer pointing at complex referents, even if it requires movement, over verbal descriptions \cite{Bovo2022}.The importance of pointing in collaborative VR environments is emphasized by research focused on enhancing pointing gestures accuracy. Techniques like warping or distorting deictic gestures have been proposed to improve collaboration \cite{Sousa2019}. \citeauthor{Mayer2018TheEnvironments} explored offset correction and cursor effects on mid-air pointing, finding that subtle redirection of a user's arm to align with their gaze can significantly improve pointing accuracy from an observer's perspective \cite{Mayer2018TheEnvironments}.

This collection of research emphasises that when the gold standard for immersive multi-modal dialogue comprehension  (i.e., the human) faces uncertainty regarding referent clarity, it consistently resorts to \textit{pointing} for disambiguation, whether in the speaker or observer scenario. Therefore, our work uses pointing behaviour to enhance machine comprehension of immersive human-human dialogues.

\subsection{Non Verbal and Multimodal Interaction}
\label{rw:non_verbal-interaction}

Prior work in HCI leverages natural nonverbal communication, such as deictic pointing gestures or interpreting visual attention concurrently with speech commands. 
In both cases, speech+pointing or speech+gaze, the use of visual attention or pointing gestures aid the process of understanding the referent intended by the user. \citeauthor{10.1145/800250.807503}'s seminal work "Put-That-There" \cite{10.1145/800250.807503} explored the integration of voice commands and hand gestures to enhance user interaction within graphical interfaces. Similarly \citeauthor{Miniotas2006SpeechGaze} studied the integration of eye gaze with speech, especially for interactions with small, closely spaced on-screen targets \cite{Miniotas2006SpeechGaze}. Recent works have pivoted towards understanding the visual context and harnessing natural collaborative communication for enhanced interactions. Specifically, \citeauthor{Bovo2023Speech-AugmentedAnalysis} explored how head direction serves as an indicator of visual attention and speech \cite{Bovo2023Speech-AugmentedAnalysis}. 


Similarly, \citeauthor{Mayer2020EnhancingWorldGaze} \cite{Mayer2020EnhancingWorldGaze} enhanced mobile voice assistants' understanding of nearby buildings by incorporating GPS location and user's head gaze (i.e. user's location and visual direction). \citeauthor{romaniak2020nimble} \cite{romaniak2020nimble} introduced 'Nimble,' a mobile interface combining visual question-answering models with gesture recognition for more intuitive user interactions. Similarly, GazePointAR, a wearable AR system, resolves speech query ambiguities using eye gaze, pointing gestures, and Human-AI conversation history \cite{lee2024gazepointar}. \citeauthor{penzkofer2021conan} explored collaborative behavior in multimodal conversations to extract metrics like collaboration quality and the impact of technology usage \cite{penzkofer2021conan}. However, none of these prior works explored the collaborative patterns of visual attention or pointing behaviour specifically towards the problem of identifying a referent of a REs; instead, they focus on interactions rooted in single-user nonverbal communication and single-event interactions. In our work, we leverage similar information for coreference resolution in recorded speech transcripts of immersive multimodal conversations. 

%% file: 3-system_design_reduced.tex
\section{System and Implementation}




Our system processes a VR recording of an immersive spatial conversation to identify implicit referring expressions (RE) and perform coreference resolution for each. The input includes the VR session's audio, eye-gaze, and laser pointer data. The output is an explicit referent for each spatial RE identified.

The process involves multiple steps: converting speech into a diarized transcript, detecting implicit spatial REs, and analyzing non-verbal cues (gazing and pointing) to identify the object of interest in a 3D scene for each RE. The system uses the transcript, non-verbal cues, and visual scene metadata (objects' names) to generate an augmented transcript. Finally, coreference resolution is performed for each spatial RE.

Implemented on an Intel Core i7 with 16GB RAM, the system processes each 15-minute session in about 5 minutes, including multimodal data processing and transcript generation.

\subsection{Core Concepts}
To contextualize our system's design, we first define the core linguistic concepts it is built to address. The primary challenge is \textbf{coreference resolution}: the task of associating ambiguous references, such as pronouns like ``it'' or ``this,'' with the specific entities they refer to. These ambiguous references are a type of \textbf{Referring Expression (RE)}, which is any phrase used by a speaker to identify an entity in their environment. Our system specifically targets \textbf{implicit spatial REs}, where the referent is not explicitly named within the phrase (e.g., ``I like this''), as opposed to \textbf{explicit spatial REs} that clearly name the referent (e.g., ``I like this \textit{mirror}''). Implicit REs introduce two distinct challenges that guide our system's design. The first is \textbf{exophora}, where the referent is completely absent from the dialogue and can only be identified through non-verbal cues and the shared visual context. The second, more frequent, challenge is \textbf{endophora}, where the referent is named elsewhere in the transcript, creating textual ambiguity between potential candidates mentioned earlier (\textbf{anaphora}) or later (\textbf{cataphora}) in the conversation. Our system is therefore designed to resolve both exophoric and endophoric ambiguity by augmenting the transcript with non-verbal data, providing the necessary context for accurate coreference resolution.

\begin{figure*}[t]
\centering
\includegraphics[width=1\textwidth]{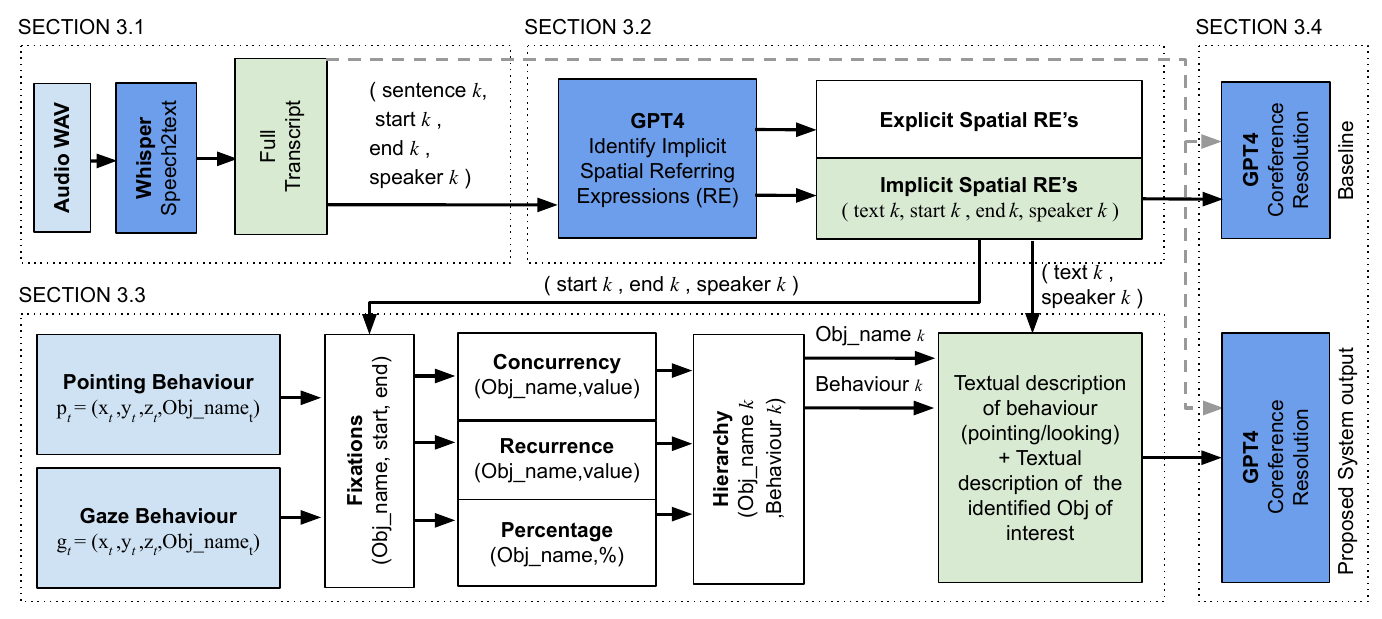}
\caption{The system architecture, as depicted in the diagram, consists of four main components: Transcript Generation (Section~\ref{system:Transcripts}), Spatial Referring Expression (RE) Identification (Section~\ref{system:IdentifyAllRef}), Object of Interest Identification (Section~\ref{system:IdentifyObjectOfInterest}), and Coreference Resolution (Section~\ref{system:CoreferenceResolution}). In Transcript Generation (Section~\ref{system:Transcripts}), the Whisper AI model transcribes audio with time-stamping. Spatial RE Identification (Section~\ref{system:IdentifyAllRef}) uses GPT-4 to detect Implicit Spatial REs for coreference resolution, with performance detailed in Fig~\ref{fig:F1score}(b). Object of Interest Identification (Section~\ref{system:IdentifyObjectOfInterest}) analyzes pointing and gaze to identify objects in Implicit REs, employing fixation calculation, concurrence recurrence, and a hierarchical selection method; results shown in Fig~\ref{fig:ObjectOfInterest}, and the fourth step generates descriptions of non-verbal behaviours and identified objects. Finally, Coreference Resolution (Section~\ref{system:CoreferenceResolution}) contrasts a baseline system (which uses only the speech transcript) with our proposed system (which integrates the transcript with the non-verbal behaviour and the object that it entails). }
\label{fig:system}
\end{figure*}

\subsection{Transcript} 
\label{system:Transcripts} 
We utilize the timestamped Whisper AI model \footnote{\url{https://github.com/linto-ai/whisper-timestamped}} to transcribe VR session recordings. Each participant's audio track is transcribed separately, preserving speaker identity and enabling diarization. The individual transcriptions are then merged, appending speaker identifiers to each segment and arranging them chronologically. The resulting transcript of the collaborative communication includes temporal timestamps for each word and sentence, along with speaker identity information.

\subsection{Identify Implicit Spatial Referring Expressions (RE)}
\label{system:ImplicitSpatialRefrences}
Implicit spatial referring expressions (REs) reference a location or spatial relationship indirectly without explicitly naming the location or object. For instance, "right on top of it!" is an implicit spatial RE, whereas "right on top of the couch!" is an explicit spatial RE. To identify these, we first \textit{identify all spatial REs} and then \textit{classify each as either implicit or explicit}.

\subsubsection{Identify Spatial referring expressions}
\label{system:IdentifyAllRef}


GPT-4 is used to identify spatial referring expressions (REs) in the VR transcript. The prompt defines the system's role: \textit{"you are a system that identifies spatial referring expressions related to objects/places in a given sentence"}. The prompt also specifies the response format: \textit{"list them in the following JSON format:\{ "spatial\_referring\_expressions": ["referring\_expression",...]\}"}. We further refine the prompt to clarify what constitutes a spatial RE to an object or place and what does not. We exclude REs where the referent is a person (e.g., you, me, we, guests) or temporal REs (e.g., now, then, today, tomorrow). Additionally, we specify not to list REs related to objects not currently present in the scene (e.g., "there is no oven") or abstract/metaphorical entities.


Next, we iterate through each sentence in the transcript. For each, we send the refined prompt to GPT-4. The model's responses (i.e., identified REs) are saved within the JSON file of the transcript. The full prompt can be seen in the supplemental material 5.3, Listing 1.

\subsubsection{Classify implicit and explicit referring expression (RE)}
\label{system:classifyEachRef}
After identifying each spatial RE, the system further classifies them to identify those requiring coreference resolution (i.e., implicit spatial REs). We generate a prompt for GPT-4 containing the spatial RE and its sentence, asking the system to determine if the RE's referent noun is present within the sentence. If a spatial RE's referent is contained within the sentence, we classify it as \textit{explicit}; if not, we classify it as \textit{implicit}. The full request sent to the API can be seen in the supplemental material 5.3, Listing 2.

\subsection{Identify Object of Interest}
\label{system:IdentifyObjectOfInterest}

To pinpoint the object of interest within a scene, we analyze the spatial behaviors of people, using time series data for gaze and pointing actions. Each sample includes a 3D vector \texttt{(x, y, z)} representing the gaze position or pointing direction, along with the name of the intersected object in the 3D model. We process this data by identifying the \textit{gaze and pointing fixations} on objects within the 3D scene, calculating \textit{concurrent}, \textit{recurrent}, and \textit{individual} behaviors of pointing and gazing at objects, and finally \textit{prioritizing the identified objects hierarchically} based on these behaviors.

\subsubsection{Gaze and Pointing Fixations}
\label{system:fixations}


Fixations refer to periods during which the person's attention (eye gaze or laser pointer) remains steady on a specific point in space. By analyzing fixations, we can discern meaningful samples within a signal—such as those indicating the person is pointing at or looking at an object—from noise, like when the person is merely glancing around the room.



We calculate fixations using the I-DT (Dispersion-Threshold Identification) algorithm \cite{SalvucciIDT}, extended to use additional scene information. Our gaze samples, recorded as points \texttt{(x, y, z)} in a virtual reality (VR) environment, are computed by casting a ray from the eye position along the eye tracker's recorded direction. This includes identifying the geometry hit by the ray in the VR environment. This helps determine whether the person is fixating on an object or moving toward a new target. If the pointer moves to a different object, the fixation is either completed or reset.



When using a laser pointer in a virtual environment, people often initially point incorrectly and then adjust. To address this, we calculate both eye and 'laser fixations'—periods where the laser pointer is steadily aimed at one location. This filters out data related to adjustments, keeping only the informative parts where the laser is fixed on the intended object. By considering both eye and laser fixations, we more accurately determine the person's attention and interaction within the VR environment. We use a 0.5-degree threshold and a 100-millisecond duration threshold as proposed by \citeauthor{SalvucciIDT} \cite{SalvucciIDT}.

\subsubsection{Fixation Concurrency over object}
\label{system:Concurrency} 
Previous research highlights that synchronizing verbal communication with visual attention reduces misunderstandings during collaboration \cite{Vrzakova2019}. Building on this, we believe that measuring simultaneous visual attention or pointing provides better insights into conversation focus than analyzing individual behavior alone.

Our algorithm detects overlapping fixations on the same object by two individuals. Each fixation is represented as a tuple with the object and start and end times. We measure concurrency by counting how often both individuals fixate on the same object at overlapping times, normalizing this count by the maximum possible concurrent fixations. This approach quantifies the extent of shared attention on objects during collaborative interactions.





\subsubsection{Fixation Recurrence over object}
\label{system:Recurrence}

A variation of concurrent attention is recurrent fixation, where two people focus on the same object in a 3D scene at different times. One person makes a comment and fixates on an object momentarily before looking away. The second person, guided by cues, then shifts their attention to the same object later. Although not simultaneous, this shows synchronization of verbal communication with visual attention.

To detect recurrent fixations between two people, we use their fixations to compute the recurrence of an object. The recurrence value is the total duration of fixations on the object by both individuals, divided by twice the total time of the REs. This approach measures how often both people direct their attention to the same object, indicating shared focus. Recurrent fixations are calculated for both gaze and pointing.





\subsubsection{Individual fixations over object}
\label{system:Individual}



We also calculate the percentage of time an object \( o \) was fixated on. This percentage is determined by dividing the total duration of fixations on the object \( D(o) \) by the total duration of all fixations \( T \), and then multiplying by 100. This gives us the proportion of the experiment time that each object held the participants' attention.

\subsubsection{Hierarchical Selection}
\label{system:Hierarchy}


Since LLMs are primarily designed to predict and generate human language based on probabilistic models, their handling of numbers \cite{wallace2019nlp,lu2022survey} is not as precise or reliable as their processing of textual information. Consequently, they are less effective at handling numerical data, such as behavioural signals (i.e., non-verbal cues). To address this limitation, we supply the LLMs with deterministic answers instead of numerical quantities by developing a Hierarchical Selection algorithm. This algorithm establishes a hierarchy of importance for the metrics identified previously, using it as a fallback mechanism if a behaviour measure combination is not occurring.


Pointing, a deliberate action requiring effort, strongly indicates intention and attention, while gazing is more reflexive and influenced by various factors. Thus, pointing behaviour is prioritized over gazing. Additionally, the context of these behaviours is crucial; synergistic behaviours (concurrent or recurrent pointing/gazing) are more informative than individual behaviours, indicating shared focus and likely discussion topics.


This hierarchy guides the identification of the object of interest. Synergistic pointing behaviour is prioritized, selecting objects pointed at concurrently or recurrently for the most time. If absent, individual pointing by the speaker of the implicit spatial reference is considered. If no individual pointing occurs, visual attention (gazing) is analyzed, prioritizing objects gazed at concurrently or recurrently, followed by objects the speaker gazed at the most.


The output includes the selected behaviour measure (concurrent/recurrent/individual with gaze/pointing), the object with the highest percentage of fixation for that measure, and the person performing the implicit spatial RE.

\subsubsection{Generate a description of the object of interest and the behaviour used to determine it}
\label{system:GenerateTextDescription}

To make the implicit spatial RE less ambiguous and, therefore, facilitate coreference resolution, we integrate the object identified in the previous step into the transcript in textual form, forming an \textit{augmented transcript}. The previous steps provided details such as the name of the recognised object, the behaviour measure leading to its identification, and the person performing the implicit spatial RE. Using this information we enhance the transcript by adding contextual information, such as "Person X was pointing/looking at Object Y" or "Both people concurrently observed Object Z" to the end of the sentences containing the implicit spatial RE. For example, a sentence in the augmented transcript might appear as: "[01:00] P1: This looks weird. [P1 was pointing at the sofa]."

\subsection{Coreference Resolution}
\label{system:CoreferenceResolution}

Coreference resolution in GPT-4 uses two methods: the baseline (speech-only transcript) and our proposed system (augmented transcript). Initially, a system description defines GPT-4's role in resolving implicit spatial references (REs) within sentences. The full transcript is included for each coreference resolution attempt. GPT-4 processes up to 8,192 tokens, sufficient for our transcripts averaging 1,841.625 tokens (max 3,066, min 1,001). Each implicit spatial RE is addressed individually within the transcript, updating the prompt with the specific RE sentence. (See supplemental material 5.3, Listing 3 for the complete prompt.)

%% file: 4-data_collection_reduced.tex
\section{Data collection for System Evaluation}
There are currently no existing datasets that encompass collaborative speech, eye-gaze, and contextual information. Therefore, to assess the effectiveness of our proposed system, we compiled a dataset by recording 6 pairs of participants who were asked to collaboratively review a virtual apartment~\footnotemark[1]  ~\footnotemark[2]  scene in VR (Figure 3) capturing their speech and eye-gaze within the 3D space. This study received review and approval from our institution's internal ethics review process.

\subsection{Apparatus}
\label{subsec:apparatus}
Each participant used an Oculus Quest Pro, which rendered the scene and collected audio, gaze, and gesture data. Audio was recorded via an internal microphone, gestures via controllers, and gaze via built-in cameras at 120Hz with 0.5 degrees accuracy. We developed a custom application using Unity 2022.3.2f1 and the Oculus Unity SDK. This application rendered a 3D scene featuring two apartments in a real-time session where participants' avatars, represented using the Oculus Avatar SDK, interacted. Participants navigated using thumbstick controllers and used a laser pointer tool activated via controller buttons. The embedded microphone and speakers enabled verbal communication between participants. Avatar movements, including head and hand gestures, were streamed with low latency using the Photon Fusion v2 Network API to synchronize their positions and behaviors across all participants' scenes. A moderator could communicate audibly but was not visually rendered within the VR session.

\subsection{Participants}
\label{subsec:participants}
We recruited 12 participants (4 women, 8 men) with  the following inclusion criteria: being a fluent English speaker and having normal, or corrected-to-normal vision. Participants received compensation of \$75CAD, and sessions lasted approximately 60 minutes.

\begin{figure*}[t]
    \centering
    \includegraphics[width=\linewidth]{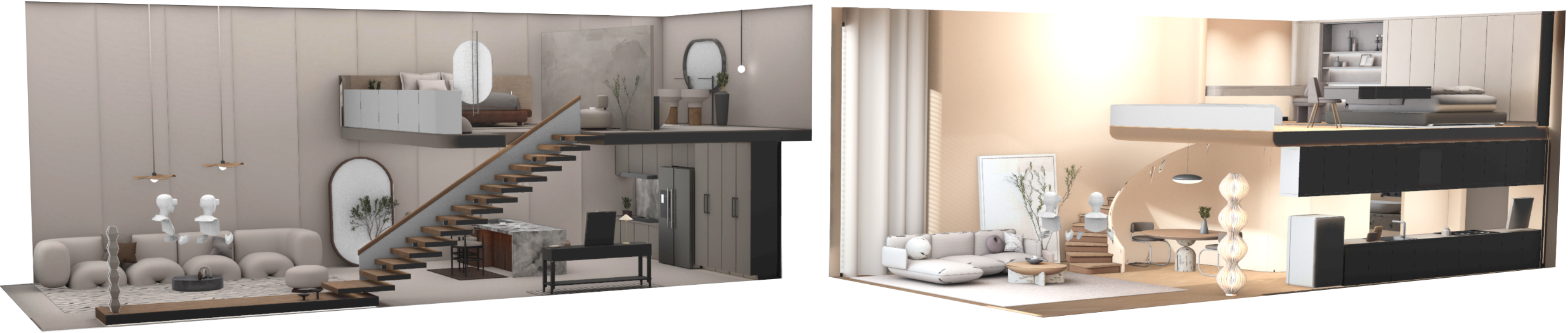}
    \caption{The two apartment scenes reviewed by participants. Participants were asked to collaboratively review these two virtual apartments in VR. \protect\footnotemark[1] \protect\footnotemark[2]}
    \label{fig:3Denv}
\end{figure*}

\footnotetext[1]{\url{https://sketchfab.com/3d-models/vr-loft-living-room-baked-f3e6f16527af4465858a34cc1e9e7a2b}}
\footnotetext[2]{\url{https://sketchfab.com/3d-models/vr-morden-loft-apartment-baked-dd252381b69d41f883083677e56a7f3e}}

\subsection{Task}
\label{subsec:task}

Participants engaged in an open-ended collaborative task involving navigating a virtual apartment scene, where they identified and discussed design aspects such as issues, considerations, and personal preferences. This task draws inspiration from recent VR applications in architectural reviews, as evidenced by studies \cite{liu2020evaluating,liu2014virtual,lopez2018virtual} and commercial uses \cite{thewild,workshopxr,arkio} aimed at pre-construction evaluation of architectural designs.
Experimenters guided participants encouraging them to examine spaces, furnishings, fixtures, and equipment, and to discuss observations with their collaborator. Participants had the freedom to comment on any design aspect without specific requirements. It was emphasized that consensus was not necessary, there were no right or wrong solutions, and the main goal was to engage in an environment-focused discussion. Supplemental material includes five conversation excerpts with labeling, co-reference resolution, and associated videos.

\subsection{Procedure}
\label{subsec:procedure}

Each data collection session followed a structured procedure. Initially, participants were presented with an informed consent form detailing the study's purpose, participant expectations, and data handling procedures, which they read and signed. Participants were informed that they would navigate a virtual 3D model of an apartment, collaborating to identify and discuss design elements. Next, participants received an orientation to the head-mounted display (HMD), focusing on proper fitting and adjustment of the HMD. They also received instruction on using the device's controllers for navigation and interaction within the application. Participants then engaged in the main task, which lasted approximately 15 minutes. Additionally, they performed an unrelated VR task for approximately 15 minutes per apartment. After completing both tasks, participants underwent a comprehensive debriefing session where the study's purpose and data collection rationale were explained.

\begin{figure*}[t]
    \centering
    \noindent
    \begin{minipage}{0.32\linewidth}
        \includegraphics[width=\linewidth]{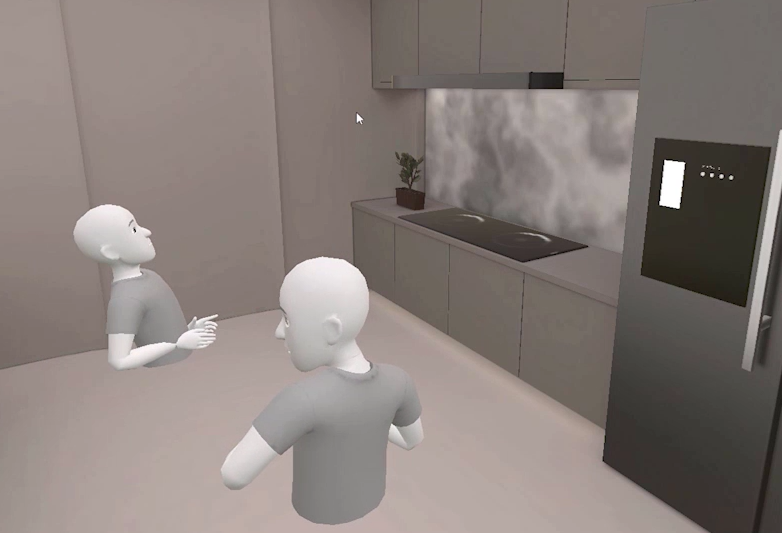}
    \end{minipage}
    \hfill
    \begin{minipage}{0.63\linewidth}
    \begin{Verbatim}[commandchars=\\\{\}]
[02:34 --> 02:36] u7 : There are no handles on the cabinets.
[02:37 --> 02:38] u7 : Or maybe these are like push ones.
[02:39 --> 02:42] u7 : And...
[02:44 --> 02:46] u8 : But it's a good design overall.
[02:47 --> 02:51] u7 : I know. Like I got a lot of time
                         picking on it .
[02:52 --> 02:53] u8 : Maybe if you go upstairs, we'll find...
    \end{Verbatim}
    \end{minipage}
\caption{This snapshot shows participant's in the environment while the transcript captures the dialogue.}
    \label{fig:cabinet-discussion}
\end{figure*}

\begin{figure*}[t]
    \centering
    \begin{Verbatim}[commandchars=\\\{\}]

| \exophora{implicit RE Exophora} | \REendophora{implicit RE Endophora} | \spatialexplicitRE{explicit RE} | \generalRE{non sptial RE} 
| ( referent ) -->  resolved by human labeler

[02:34 --> 02:36] u7 :   There are no handles on \spatialexplicitRE{the cabinets}\textsuperscript{1}.
[02:37 --> 02:38] u7 :   Or maybe \REendophora{these (cabinets)}\textsuperscript{2} are like push ones.
[02:39 --> 02:42] u7 :   And...
[02:44 --> 02:46] u8 :   But \exophora{it's (kitchen design)}\textsuperscript{3} a good design overall.
[02:47 --> 02:51] u7 :   I know. Like I got a lot of time picking on \exophora{it (kitchen design)}\textsuperscript{4}.
[02:52 --> 02:53] u8 :   Maybe if you go upstairs, we'll find...
    \end{Verbatim}
\caption{An example of the manual annotation process. The transcript corresponds to the snapshot in \autoref{fig:cabinet-discussion} and the manual labels constitutes the ground truth used to test the system.}
    \label{fig:manual-labeling}
\end{figure*}
\begin{figure*}[t]
    \centering
    \begin{Verbatim}[commandchars=\\\{\}]

| \correct{correctly identified implicit spatial RE}  | \wrong{incorrectly identified implicit spatial RE}    
| \argumented{augmented transcrip} | \correctt{correctly resolved}  | \wrongg{incorrectly resolved} 

[02:34 --> 02:36] u7 :   There are no handles on the cabinets\textsuperscript{1}.
[02:37 --> 02:38] u7 :   Or maybe \correct{these}\textsuperscript{2} are like push ones.
                       \argumented{[u7 and u8 concurrently  looking at the kitchen cabinets]}
                       these\textsuperscript{2} --> baseline:  \correctt{cabinets}, system: \correctt{ kitchen cabinets}
                       
[02:39 --> 02:42] u7 :   And...
[02:44 --> 02:46] u8 :   But \correct{it's }\textsuperscript{3} a good design overall.
                       \argumented{[u8 looking at the kitchen cabinets]}
                       it's \textsuperscript{3} --> baseline:  \wrongg{design overall}, system: \correctt{ kitchen overall design}
[02:47 --> 02:51] u7 :   I know. Like I got a lot of time picking on \correct{it}\textsuperscript{4}.
                       \argumented{[u8 looking at the lamp]}
                       it \textsuperscript{4} --> baseline:  \wrongg{the apartment}, system: \wrongg{ the lamp}
[02:52 --> 02:53] u8 :   Maybe if you go upstairs, we'll find...
    \end{Verbatim}
    \caption{An example of the system’s output showing the transcript augmentation corresponding to the snapshot in \autoref{fig:cabinet-discussion}, together with the final coreference resolution results for both the baseline and the proposed system. }
    \label{fig:system-output-example}
\end{figure*}

\subsection{Collected Data}
Various data were collected during the task, all recorded at 120Hz. Eye-tracking data was collected recording the x, y, and z coordinates of gaze locations on the scene's objects. Laser pointing data was similarly collected, recording the x, y, and z coordinates of the laser's location on the scene's objects. Additionally, participant conversations were recorded. All participant data, was recorded by the monitoring application. Data synchronization and time stamping were ensured using a common time reference. An example snapshot of the collected data is shown in Figure~\ref{fig:cabinet-discussion}, which displays a view from a participant's perspective alongside a segment of the transcribed conversation.

\subsection{Data Labelling}

Three of the authors manually labeled transcribed audio recordings to identify spatial referring expressions (REs) and classify them as implicit or explicit spatial REs. Implicit REs were further categorized based on whether they referred to an entity present in the transcript (endophora) or one absent, relying on visual context and non-verbal communication (exophora). Subsequently, we determined the referent and corresponding target geometry in the 3D scene for all references. The manual labeling process was carried out in order to identify the ground truth and compare the results of our proposed system. Given that three distinct raters participated in this labeling task, we assessed inter-rater reliability to ensure consistency, as discussed in Section~\ref{InterRater}. 
An example of labeled data is shown in Figure~\ref{fig:manual-labeling}, which displays implicit RE (exophora/endophora) and explicit RE. Further examples of labeled data are available in the supplemental material 1--4.

\subsubsection{Labeling Spatial Referring Expressions} \label{SpatialRef} Manual labeling begins by identifying all expressions and phrases which reference objects or places in the scene. These phrases or expressions might contain demonstrative deixis "this chair" or "the other one", adverbs of place "it would be better there" and prepositional phrases "...under the counter-top". We exclude certain RE from the process: objects not currently present (e.g., 'there is no oven'), temporal deixis (e.g., 'in the room we were before'), and abstract or metaphorical entities (i.e. in the dialogue `P1: "We should isolate this shower."' `P2: "I'd like that!"' in the second sentence, P2 refers to the idea of isolating the shower, not the shower itself).

\subsubsection{Labeling Implicit, Explicit, Endophora, Exophora} \label{ImplicitReferences} Once all the spatial referring expressions are identified, we categorize each of them as either implicit or explicit. For example, an explicit spatial reference would be "this chair" as it clearly identifies a specific object in the scene. An implicit spatial reference might be "it looks weird," because "it" might refer to an object previously mentioned or an object that the participant is pointing at in the VR environment, without explicitly naming the object. For each implicit RE it is also labeled based on its reference type: whether it alludes to an entity explicitly named within the transcript (endophora) or to an entity not mentioned in the transcript (exophora).

\subsubsection{Labeling Resolved Referent} \label{intendedTarget} For all the labeled implicit RE, we then annotate their intended target by watching the video corresponding to when the reference was made. The video provides context for where the users were and what they were watching, as well as understanding if they were performing pointing gestures, in order to understand what participants were referring to.

\subsubsection{Labeling Target Geometry} \label{TargetGeometry} Once all the implicit references' target objects are identified, we then annotate the corresponding geometry name in the 3D scene by manually inspecting the geometry.

\subsubsection{Inter Rater Reliability} \label{InterRater} 
All labeling was conducted by authors, where each label was initially assigned by a Labeler and subsequently reviewed by a Rater. The Rater either agreed or disagreed with the Labeler's assignment, adding additional labels in cases of disagreement. Disagreements were documented with comments for each specific label and resolved through discussion at a later stage. To evaluate the consistency of the labeling process, we performed an inter-rater reliability analysis, initially resulting in an agreement percentage of 80.7\%. Following discussion and resolution of disagreements, a final complete agreement on all labels was achieved among all labellers.

%% file: 5-result-reduced.tex
\section{Results}
\label{results}

We  evaluate our system's ability to identify implicit spatial REs (Figure~\ref{fig:F1score}d), to identify the specific geometry referred to (Section~\ref{result:Identify Object of Interest}), and to perform coreference resolution for the identified implicit spatial REs (Section~\ref{result:CoRes}). 
The described manual labeling procedure provide the ground truth for the evaluation.  Performance measures of precision and F1-score are calculated for each individual user. Data aggregation is done per user rather than per conversation, considering each user's unique communication style, preference in using non-verbal communication, and vocabulary. Because the collected data does not conform to a normal distribution, we use a nonparametric bootstrap approach with 1,000 resamples to estimate the sampling distribution and derive 95\% confidence intervals for evaluating differences.

\subsection{Identify Implicit Spatial RE}
\label{result:CoRes}
To evaluate the performance of the implicit spatial RE identification phase, the F1-score was calculated (Figure~\ref{fig:F1score} (d)). Out of the 350 Implicit REs we manually labelled, our system correctly identified 318 and misclassified 82. Therefore when interpreting this F1-score it is important to take into account that errors generated in the identification phase directly propagate to the coreference resolution phase. 

\begin{figure*}[t]
    \centering
        \includegraphics[width=\linewidth]{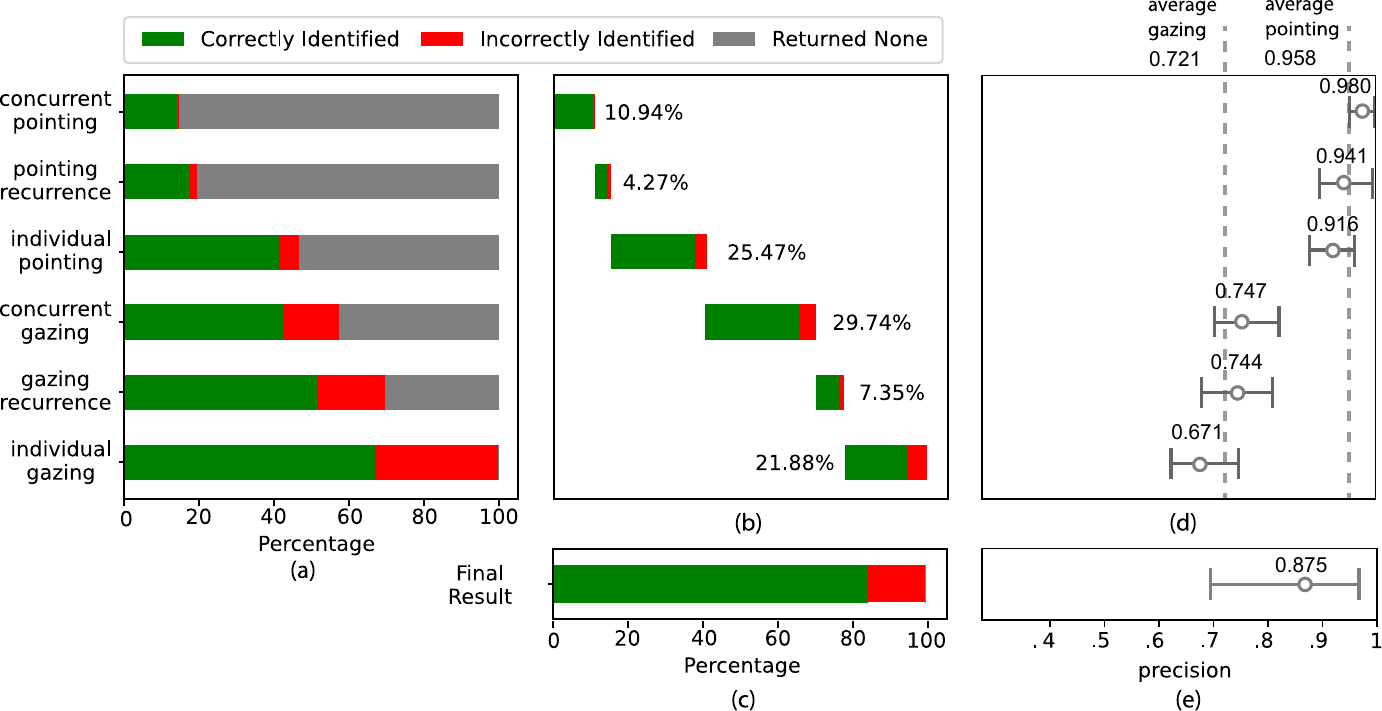}
        \caption{Identified Object of Interest:  
        (a) Plot depicting the performance of each behaviour combination in terms of identifying the correct object, the incorrect object, or not identifying an object at all. The percentages for correct, incorrect, or none are calculated for each implicit spatial RE (y-axis: behaviour measures , x-axis: percentage of correct/incorrect/none).
        (b) Plot showcasing the frequency with which each behaviour measure was selected during our hierarchical selection process. Each behaviour's percentage bar is divided into 'correct' and 'incorrect'. The scenario of not identifying anything is excluded since this condition triggers our hierarchical selection process to fallback onto the subsequent behaviour measure(y-axis: behaviour measures , x-axis: percentage of correct/incorrect).
        (c) Plot presenting the percentages for correct,incorrect and none as results of the hierarchical selection process (y-axis: behaviour measures , x-axis: percentage of correct/incorrect).
        (d) Plot displaying the object identification precision, defined as the ratio of correctly predicted positives to total predicted positives, for each behaviour measure (y-axis: behaviour measures , x-axis: precision on a unit scale).
        (e) Plot highlighting the final precision of our object identification process. }
        \label{fig:ObjectOfInterest}
\end{figure*}

\subsection{Identifying Object of Interest}
\label{result:Identify Object of Interest}

We evaluate the effectiveness of different behaviour measures, such as concurrent pointing, pointing recurrence, individual pointing, concurrent gazing, gazing recurrence, and individual gazing, in identifying the object of interest for each implicit RE. Since no established techniques utilize collaborative speech, gaze, and world semantics, direct comparisons with other methods are not feasible. Therefore, we evaluate the merit of each behavior combination in detecting the object of interest through an ablation study. This study assesses the contribution of each individual component to the overall effectiveness of our system (Figure~\ref{fig:ObjectOfInterest} (a)). We do so by comparing the results with the ground truth to determine how many correct and incorrect objects each behaviour measure returns (Figure~\ref{fig:ObjectOfInterest} (a,d)). While we compare all behaviours with one another, our system then selects a single one via the hierarchical selection mechanism described previously(section \ref{system:Hierarchy}). We also calculate how many correct and incorrect objects our overall system identifies (Figure~\ref{fig:ObjectOfInterest} (b,c,e)). Using the same ground truth as a reference, we assess our system's capability in recognising the object of interest for the implicit REs.

\subsection{Effectiveness of Behaviours in Identifying Object of Interest}

To better understand the impact of the various non-verbal behaviours (gaze, pointing), we compute the number of correctly identified objects for each behaviour, as well as how frequently a behaviour did not return any result because it did not happen in tandem with the RE (see Figure~\ref{fig:ObjectOfInterest}). The data reveals that pointing behaviours, with \texttt{concurrent pointing}, \texttt{recurrent pointing}, and \texttt{individual pointing} occurrences at 15\%, 19.3\%, and 46.4\% respectively, are less frequent than gaze behaviour, which occurs at 56.8\% for \texttt{concurrent gazing}, 69.3\% for \texttt{gazing recurrence}, and 99.4\% for \texttt{individual gazing} instances. Note that individual gazing is 99.4\% rather than 100\% due to low eye tracking confidence or blinking. Furthermore, both concurrent and recurrent behaviours manifest less often than their corresponding individual behaviours (see Figure~\ref{fig:ObjectOfInterest}). For each combination, we computed precision by dividing the number of correctly identified objects by the total number of returned objects. The final precision of our system in identifying the object of interest is 0.875. The results highlight that \texttt{pointing} is more precise than \texttt{gazing}, showing a .16 increment in the precision towards identifying the object of interest (Figure~\ref{fig:ObjectOfInterest} (d)). There is a large difference between gazing and pointing (\texttt{gazing}: 95\% CI [0.667, 0.766]; \texttt{pointing}: 95\% CI [0.926, 0.980]), when comparing their precision.A comparison of \texttt{concurrent pointing} (Mean = 0.988, 95\% CI [0.963,1.000]) and \texttt{individual pointing} (Mean = 0.926, 95\% CI [0.881,0.967]) also indicates a strong difference in precision. Additionally, we assess how each combination influenced our final result through its selection in our hierarchical process (Section\ref{system:Hierarchy}). This is achieved by determining the frequency with which each behavior is chosen by our hierarchical selection process. Such frequency highlights the contribution of each behavior measure towards the goal of determining the object of interest. The results indicate that the combination contributing most to identifying the object of interest is \texttt{concurrent gazing} (representing concurrent visual attention) at 29.74\%. This is followed by \texttt{individual pointing} at 25.47\% and \texttt{individual gazing} at 21.88\%. Moreover, it's evident that the majority of errors stem from \texttt{individual gazing} at 5.12\% and \texttt{concurrent gazing} at 4.44\%, with \texttt{individual pointing} contributing 3.07\% of errors.

\begin{figure*}[t]
    \centering
        \includegraphics[width=\linewidth]{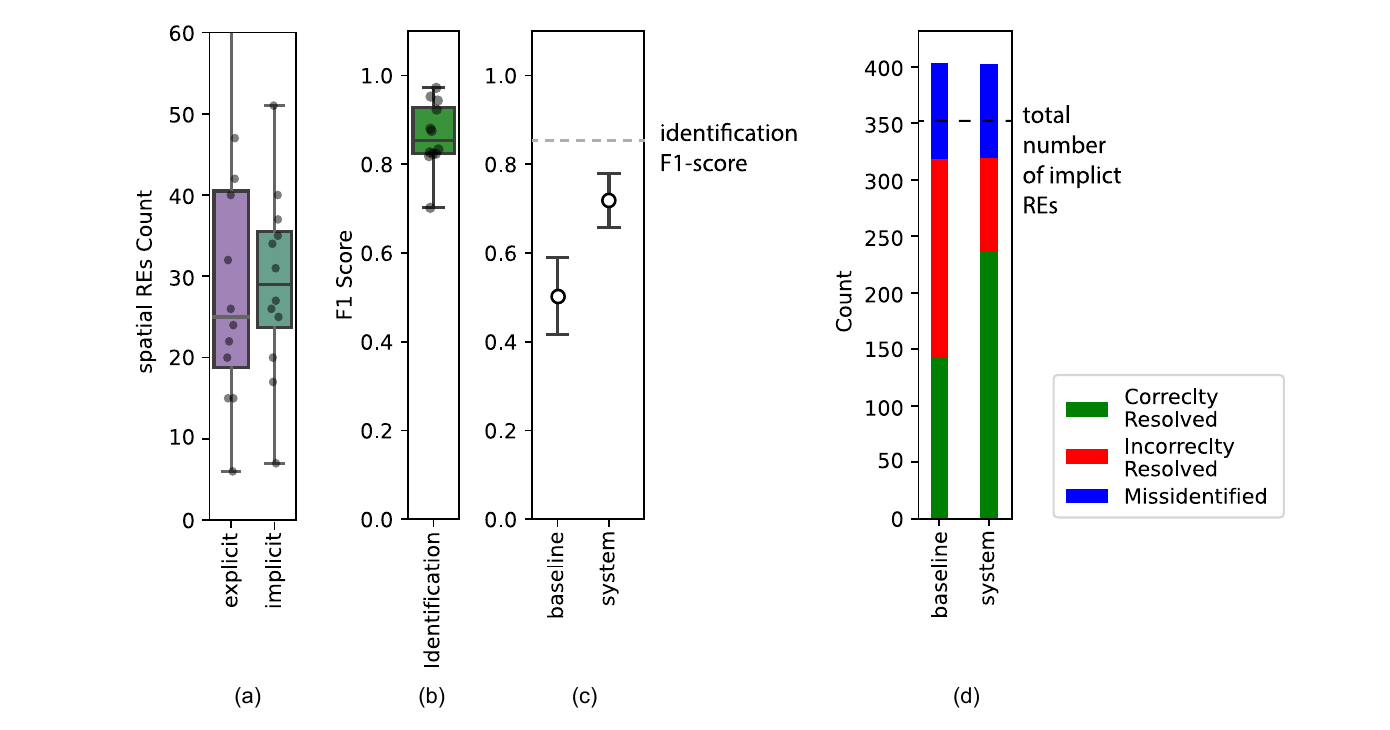}
        \caption{
            (a) Plot depicting the count of explicit and implicit REs labelled for each participant (y-axis: count per participant, x-axis: implicit/explicit).
            (b) Plot presenting the f1-score for identifying implicit spatial REs.
            (c) Plot comparing the f1-score for the coreference resolution task executed by GPT-4 using the speech transcript (baseline) versus our augmented transcript with metadata (y-axis: f1 score, x-axis: baseline/system).
            (d) Plot depicting the count of correctly resolved, incorrectly resolved and miss-identified (y-axis: REs count, x-axis: baseline and our proposed method).}
        \label{fig:F1score}
\end{figure*}

\subsection{Coreference Resolution}

We define a \texttt{baseline} consisting of  GPT-4 performing coreference resolution on the speech-transcript. We compare it to GPT-4 performing coreference resolution on our \texttt{system}'s augmented transcript. We compare both against the manually labelled ground truth (Section~\ref{intendedTarget}).We calculated F1-score for both \texttt{baseline} and \texttt{system} (Figure~\ref{fig:F1score} (c)). When interpreting the F1-score, it's important to take into account that errors generated in the identification phase (Figure~\ref{fig:F1score} (b)) directly propagate to the coreference resolution phase  (Figure~\ref{fig:F1score} (c)). This is because the errors between the identification system and the coreference are independent. For reference we plotted the F1-score from our implicit spatial RE's identification in Figure~\ref{fig:F1score} (c) and Figure~\ref{fig:Charactherization} (b)(d). Coreference resolution examples are available in the supplemental material 1--4.

\subsubsection{Comparing our system to the baseline. }

We compare baseline and system F1-score and we observed improved coreference resolution performance, resulting in a .21 increase in the F1 score when comparing \texttt{baseline} with a precision of .507 to \texttt{system} with a precision of .723. Comparing baseline and the system, we see a clear difference \texttt{baseline} (95\% CI: 0.507--0.584) and \texttt{system}  (95\% CI: 0.675--0.770) (Figure~\ref{fig:F1score} (c)). Results are further analyzed by categorizing them into 'Correctly Resolved', 'Incorrectly Resolved', and 'Miss-identified' (Figure~\ref{fig:F1score} (d)). The latter category includes instances where implicit REs were classified as explicit, non-REs were classified as implicit REs, or explicit REs were classified as implicit. Out of the 350 Implicit REs, there were 318 correctly identified. The baseline approach correctly resolved 142 (40.6\%), while our proposed system accurately resolved 235 (67.1\%) of them.

\subsubsection{Understanding endophora and exophora.}

\begin{figure*}[t]
    \centering
        \includegraphics[width=\linewidth]{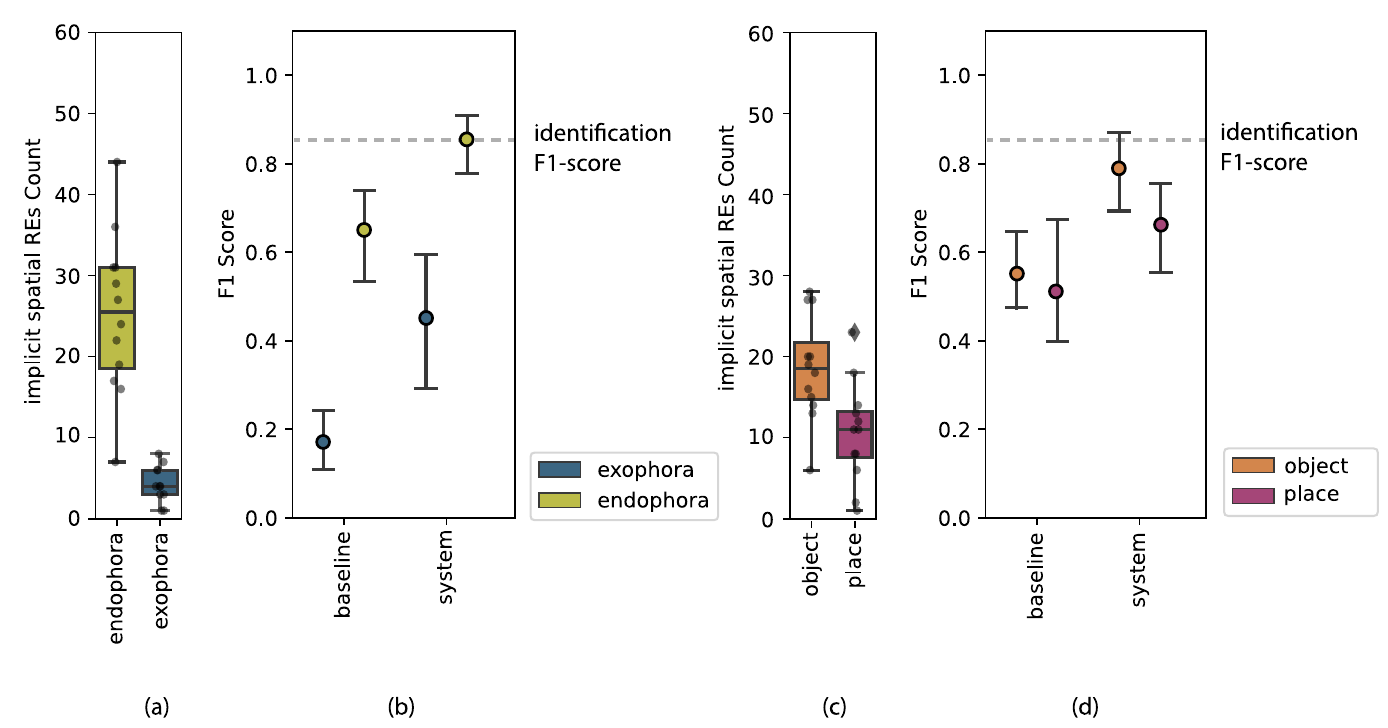}
        \caption{ (a) Plot illustrating the count of implicit REs categorized as endophora or exophora for each participant (y-axis: count per participant, x-axis: endophora/exophora).
        (b) Plot of the F1-score for the coreference resolution task by GPT-4, comparing the baseline with our system for two identified groups exophora/endophora (y-axis: f1 score, x-axis: transcript vs. transcript+metadata). 
        (c) Plot showing the count of implicit spatial REs targeting either an object or place (y-axis: count per participant, x-axis: object/place).
        (d) Plot of the F1-score for the coreference resolution task by GPT-4, comparing the baseline with our system for the two groups object/place (y-axis: f1 score, x-axis: baseline/system). }
        \label{fig:Charactherization}
\end{figure*}

We compute the performance of both the \texttt{baseline} and \texttt{system} when resolving both endophora and exophora Figure~\ref{fig:Charactherization} (a, b). 
Results show that for both \texttt{baseline} and \texttt{system}, there was an average increase of 0.44 in the F1 score for the endophora group (baseline and system combined). This effect is more pronounced in the \texttt{baseline}, which showed a 0.475 point increase in the F1 score. A comparison between the \texttt{baseline endophora} (95\% CI: 0.133--0.211) and \texttt{baseline exophora} (95\% CI: 0.544--0.696) revealed a clear difference, with the \texttt{baseline exophora} group achieving substantially higher performance. This more substantial increase aligns with expectations of \texttt{endophora} being more challenging to resolve than the \texttt{exophora}. Furthermore we observed a 0.25 increase in F1 score from \texttt{baseline exophora} to \texttt{system exophora}, (\texttt{baseline exophora}: 95\% CI 0.544--0.696; \texttt{system exophora}: 95\% CI 0.809--0.882), indicating a substantial improvement. This is consistent with expectations, as the \texttt{system} benefits from additional contextual information and enriched features that are particularly effective for resolving explicit references, such as those present in exophora scenarios. Finally the \texttt{system endophora} improved by 0.15 in F1 score over \texttt{baseline endophora}, (\texttt{baseline endophora}: 95\% CI 0.133--0.211; \texttt{system endophora}: 95\% CI 0.342--0.576), indicating a large improvement. This indicates that the supplementary information introduced by our pipeline aids the GPT-4 model in differentiating between previously mentioned entities and subsequent entities, enhancing its coreference resolution capability. Furthermore, by observing the higher number of endophora occurrences compared to exophora (Figure~\ref{fig:Charactherization} (a)), we can infer that this is where the system achieves most of its performance improvements.

\subsubsection{Understanding object and place references. }
Lastly, we calculated the F1-score based on whether the reference's target entity was labeled as an \texttt{Object} or a \texttt{Place} (Figure~\ref{fig:Charactherization} (c, d)). We observed that coreference resolution tends to be less accurate across all models when the reference's target entity is a "place." However, this difference is more pronounced in the \texttt{system} model. Specifically, \texttt{system object} exhibited a 0.18 increase in the F1 score, with moderately-overlapping confidence intervals (95\% CI: 0.717--0.861 for \texttt{system object} vs. 0.555--0.745 for \texttt{system place}). This finding aligns with our expectations, as references to \texttt{places} may lack clear geometric boundaries in the 3D scene, with identified objects (e.g., a fridge) representing only a part of the place (e.g., the kitchen) rather than encapsulating it entirely.

%% file: 6-discussion-reduced.tex
\section{Discussion}

We discuss our system's ability to identify and resolve spatial referring expressions (REs) through a transcript that has been augmented with contextual metadata about the scene via nonverbal communication (i.e., objects of interest, gaze, and pointing behaviour in relation to implicit spatial REs).

\subsection{Identifying implicit spatial referring expressions} 

Our system's first step is to identify implicit spatial referring expressions (REs) using GPT-4. Out of 350 implicit REs, the system successfully identified 318 but misclassified 63. Miss-classifications primarily involved two types: firstly, explicit REs were mistakenly classified as implicit, such as in "This is a cool little spot," where "spot" accompanied the RE, leading to incorrect classification. Secondly, expressions like "I like that." in contexts such as "There is no TV, this means digital detox." meant non-physical contexts and wrongly labeled as spatial REs. These errors suggest a need for future systems to incorporate accuracy estimates or verify if verbally expressed entities exist in the 3D environment. Several alternative methods for parsing text effectively and identifying implicit spatial REs exist, such as Stanford's CoreNLP parser \cite{manning2014_CoreNLP}. However, we chose a GPT model because our intention was to identify implicit spatial referring expressions (REs) and avoid implicit REs that refer to abstract ideas and hypothetical objects.  Given the GPT model's nuanced approach and its ability to detect subtle patterns in language a GPT model is more effective for this task compared to rule-based or traditional machine learning systems, which might overlook such patterns.  While there were several possible Large Language Models (LLM) that we could have chosen such as Llama \cite{Touvron2023LLaMA:Models}  or Falcon \cite{Penedo2023TheOnly} we used the GPT-4 API for the convenience of not having to run the model locally.  
    
\subsection{The merit of non-verbal synergies towards identifying the referent}    
Our findings underscore the value of analyzing non-verbal synergies—collaborative patterns in gaze and pointing—to resolve referents in complex human-human dialogues. This approach advances beyond prior work, which has largely focused on single-user, human-machine interactions with isolated deictic gestures \cite{romaniak2020nimble, Mayer2020EnhancingWorldGaze, Bovo2023Speech-AugmentedAnalysis, Miniotas2006SpeechGaze}. While other research has acknowledged the significance of synchronized gaze dynamics in collaborative tasks \cite{Moulder2023, Pietinen2008, Vrzakova2019, Villamor2018}, this information had not been leveraged as a direct retrieval method for identifying an object of interest. Our work expands on this by demonstrating that synergistic behaviors in both pointing (concurrent and recurrent) and gaze (concurrent and recurrent visual attention) are effective retrieval methods during collaborative discussions. Furthermore, our results revealed a critical distinction between these two modalities: pointing is a significantly more accurate predictor of the intended referent than gaze (Figure \ref{fig:ObjectOfInterest} (e)). This key finding was foundational to our system's design. We established a hierarchical selection method that prioritizes the more reliable, volitional behavior (pointing) over the more reflexive, less precise behavior (gazing). This ensures that our system leverages the most accurate non-verbal cue available to resolve ambiguity.

\subsection{Improving resolution of both endophora and exophora}    

Our system performs coreference resolution on each of the identified implicit spatial REs, resulting in an improved coreference resolution process for all implicit REs. While the results for \textit{exophora} showed significant improvements compared to the baseline, there were very few cases of them in our dataset (occurrences Figure \ref{fig:Charactherization} (a)). 
However, for those cases in the exophora group, it is evident that our system's advantage lies in incorporating novel information from the 3D scene metadata. 
The most common implicit REs were \textit{endophora} (referents present in the text but in a different sentence). Results underscore that, for the endophora group, the system's contribution is primarily in enabling the disambiguation of existing entities within the text. Therefore, our technique improves the comprehension of conversations not only by introducing new information but also by facilitating the coreference resolution process of selecting existing entities within the text through the augmented transcript.

\subsection{Extending Vision-Language Models}

Vision-language models (VLMs),  aim to understand visual context using neural networks like CNNs, MultiModal BERT, ResNet, or RetinaNet \cite{Hong20233D-LLM:Models, KongWhatCoreference, Kottur2021SIMMCConversations, Yu2019WhatDialogues, Yu2022VD-PCR:Resolution}.  These models analyze the visual scene, segment entities, and model relationships based on textual mentions. However, they can be prone to errors in scenarios where multiple objects within the user's field of view could relate to an ambiguous referring expression, such as saying "I like that cake" in a cake shop. In contrast, our approach offers an alternative method to model relationships between entities in the visual scene without relying on detailed visual analysis like VLMs. We emphasize the modeling of relationships between verbal communication and scene objects by leveraging dynamics in individual and collaborative non-verbal communication. For instance, collaborative eye-gaze and pointing can serve as additional inputs to neural networks, providing crucial information for accurately modeling relationships. This includes using eye gaze and pointing to indicate where visual attention aligns during verbal communication, akin to the temporal parsing of mouse traces demonstrated by \citeauthor{Goel2022WhoNarrations} \cite{Goel2022WhoNarrations}. This paper extends prior work on visual coreference resolution \cite{Yu2019WhatDialogues, Yu2022VD-PCR:Resolution, KongWhat2014WhatAre, Hong20233D-LLM:Models, Kottur2021SIMMCConversations, GuoGRAVL-BERT:Resolution}, highlighting novel inputs such as synergistic gaze and pointing that can operate alongside or independently of visual scene analysis, thereby contributing to higher accuracy in coreference resolution.

\subsection{Reliance on the 3D model and its granularity}

It is important to note that referents varied in granularity, as no specific constraints were imposed on the participants (i.e. to only refer to objects). When a spatial reference refers to a place (e.g., kitchen, bathroom, or other area), which is an aggregate of objects, the metadata (3d model name) might not represents the entity the speaker is referencing.  While some object names might include the location's name (e.g., "kitchen cabinets" or "bathroom sinks"), others might not (for instance, "faucet" could be located either in the kitchen or bathroom). 
Our analysis shows a lower F1 score when resolving place-related entities compared to objects. For instance, if the referent is the kitchen but the speaker is pointing the laser at the fridge, while the fridge is part of the kitchen, it is not the \textit{kitchen} itself. By describing this verbally (e.g., "[P1 pointing at the fridge]"), we do not constrain GPT-4, which, being an abstractive generative model, can infer, based on the richness of the RE and surrounding dialogue, that the referent might be the kitchen and not the fridge. Nevertheless, our results indicate that our system performs statistically worse when the referent is a place than when it is an object. We acknowledge that our 3D model could also be set up with more granular object names and using a hierarchical structure (e.g., a "faucet" would be a descendant of a "sink", which would be part of "kitchen"). Future work should explore the appropriate level of abstraction and granularity in the 3D model to address this limitation.

\subsection{Utility of coreference resolution for immersive conversations recordings}

There is a growing interest in spatial computing devices such as VR/AR for professional collaborative applications. Consequently, as user adoption increase, there will be an increasing number of conversations, such as meetings and design reviews, that will be conducted in VR and recorded for later review, summarization, and archiving. As more and more conversations get recorded, the desire to automatically process these conversations and extract salient moments will increase. However, human-human dialogues present challenges for machine comprehension for various reasons, such as determining the object being referred to using implicit referring expressions. Therefore, we argue that the impact of this work, by leveraging synergies of non-verbal cues to detect the referent of REs, will be instrumental in enhancing the accuracy and reliability of machine comprehension in human-human interactions, ultimately contributing to more effective and meaningful use of the recording capabilities of these spatial devices. 

\section{Conclusion}

To address the issue of identifying objects of interest during an immersive conversation, we developed a system that leverages transcribed text, eye tracking, and laser-pointing data to resolve coreferences. It detects implicit REs, identifies the object of attention in the scene using non-verbal cues, generates a textual description of the object of interest, and performs coreference resolution using the textual description generated. By analyzing the data collected during a 12-participant user study, we find that gaze and pointing data add value, with pointing data often providing highly precise (though infrequent) information about which objects share focus. Compared to a baseline with only speech information which resolved 142 cases, our system resolved 235 implicit REs showing an improvement of 26.5\%. 

%% file: references.bib
@inproceedings{Bovo2023Speech-AugmentedAnalysis,
    title = {{Speech-Augmented Cone-of-Vision for Exploratory Data Analysis}},
    year = {2023},
    booktitle = {Conference on Human Factors in Computing Systems - Proceedings},
    author = {Bovo, Riccardo and Giunchi, Daniele and Sidenmark, Ludwig and Newn, Joshua and Gellersen, Hans and Costanza, Enrico and Heinis, Thomas},
    month = {4},
    publisher = {Association for Computing Machinery},
    isbn = {9781450394215},
    doi = {10.1145/3544548.3581283}
}

@inproceedings{bai2021,
    title={{Joint Coreference Resolution and Character Linking for Multiparty Conversation}},
    author={Jiaxin Bai and Hongming Zhang and Yangqiu Song and Kun Xu},
    booktitle={EACL},
    year={2021},
    url={https://aclanthology.org/2021.eacl-main.43.pdf}
}

@inproceedings{Moulder2023,
   author = {Robert Moulder and Brandon Booth and Angelina Abitino and Sidney D'Mello},
   doi = {10.1145/3576050.3576113},
   isbn = {9781450398657},
   journal = {ACM International Conference Proceeding Series},
   month = {3},
   pages = {430-440},
   publisher = {Association for Computing Machinery},
   title = {{Recurrence Quantification Analysis of Eye Gaze Dynamics during Team Collaboration}},
   year = {2023},
}

@inproceedings{romaniak2020nimble,
author = {Romaniak, Yevhen and Smielova, Anastasiia and Yakishyn, Yevhenii and Dziubliuk, Valerii and Zlotnyk, Mykhailo and Viatchaninov, Oleksandr},
title = {{Nimble: Mobile Interface for a Visual Question Answering Augmented by Gestures}},
year = {2020},
isbn = {9781450375153},
publisher = {Association for Computing Machinery},
address = {New York, NY, USA},
url = {https://doi.org/10.1145/3379350.3416153},
doi = {10.1145/3379350.3416153},
booktitle = {Adjunct Proceedings of the 33rd Annual ACM Symposium on User Interface Software and Technology},
pages = {129–131},
numpages = {3},
location = {Virtual Event, USA},
series = {UIST '20 Adjunct}
}

@inproceedings{10.1145/800250.807503,
author = {Bolt, Richard A.},
title = {{“Put-That-There”: Voice and Gesture at the Graphics Interface}},
year = {1980},
isbn = {0897910214},
publisher = {Association for Computing Machinery},
address = {New York, NY, USA},
url = {https://doi.org/10.1145/800250.807503},
doi = {10.1145/800250.807503},
booktitle = {Proceedings of the 7th Annual Conference on Computer Graphics and Interactive Techniques},
pages = {262–270},
numpages = {9},
location = {Seattle, Washington, USA},
series = {SIGGRAPH '80}
}

@InProceedings{manning2014_CoreNLP,
  author    = {Manning, Christopher D. and  Surdeanu, Mihai  and  Bauer, John  and  Finkel, Jenny  and  Bethard, Steven J. and  McClosky, David},
  title     = {{The {Stanford} {CoreNLP} Natural Language Processing Toolkit}},
  booktitle = {Association for Computational Linguistics (ACL) System Demonstrations},
  year      = {2014},
  pages     = {55--60},
  url       = {http://www.aclweb.org/anthology/P/P14/P14-5010}
}

@inproceedings{SalvucciIDT,
author = {Salvucci, Dario D. and Goldberg, Joseph H.},
title = {{Identifying Fixations and Saccades in Eye-Tracking Protocols}},
year = {2000},
isbn = {1581132808},
publisher = {Association for Computing Machinery},
address = {New York, NY, USA},
url = {https://doi.org/10.1145/355017.355028},
doi = {10.1145/355017.355028},
booktitle = {Proceedings of the 2000 Symposium on Eye Tracking Research \& Applications},
pages = {71–78},
numpages = {8},
location = {Palm Beach Gardens, Florida, USA},
series = {ETRA '00}
}

@INPROCEEDINGS{KongWhat2014WhatAre,

  author={Kong, Chen and Lin, Dahua and Bansal, Mohit and Urtasun, Raquel and Fidler, Sanja},

  booktitle={2014 IEEE Conference on Computer Vision and Pattern Recognition}, 

  title={{What Are You Talking About? Text-to-Image Coreference}}, 

  year={2014},

  volume={},

  number={},

  pages={3558-3565},

  doi={10.1109/CVPR.2014.455}
}

@inproceedings{Miniotas2006SpeechGaze,
author = {Miniotas, Darius and \v{S}pakov, Oleg and Tugoy, Ivan and MacKenzie, I. Scott},
title = {{Speech-Augmented Eye Gaze Interaction with Small Closely Spaced Targets}},
year = {2006},
isbn = {1595933050},
publisher = {Association for Computing Machinery},
address = {New York, NY, USA},
url = {https://doi.org/10.1145/1117309.1117345},
doi = {10.1145/1117309.1117345},
booktitle = {Proceedings of the 2006 Symposium on Eye Tracking Research \& Applications},
pages = {67–72},
numpages = {6},
location = {San Diego, California},
series = {ETRA '06}
}

@inproceedings{Pietinen2008,
author = {Pietinen, Sami and Bednarik, Roman and Glotova, Tatiana and Tenhunen, Vesa and Tukiainen, Markku},
title = {{A Method to Study Visual Attention Aspects of Collaboration: Eye-Tracking Pair Programmers Simultaneously}},
year = {2008},
isbn = {9781595939821},
publisher = {Association for Computing Machinery},
address = {New York, NY, USA},
url = {https://doi.org/10.1145/1344471.1344480},
doi = {10.1145/1344471.1344480},
booktitle = {Proceedings of the 2008 Symposium on Eye Tracking Research \& Applications},
pages = {39–42},
numpages = {4},
location = {Savannah, Georgia},
series = {ETRA '08}
}

@inproceedings{Wong2014SupportDeicticPointing,
author = {Wong, Nelson and Gutwin, Carl},
title = {{Support for Deictic Pointing in CVEs: Still Fragmented after All These Years'}},
year = {2014},
isbn = {9781450325400},
publisher = {Association for Computing Machinery},
address = {New York, NY, USA},
url = {https://doi.org/10.1145/2531602.2531691},
doi = {10.1145/2531602.2531691},
booktitle = {Proceedings of the 17th ACM Conference on Computer Supported Cooperative Work \& Social Computing},
pages = {1377–1387},
numpages = {11},
location = {Baltimore, Maryland, USA},
series = {CSCW '14}
}

@article{Bovo2022,
    title = {{Cone of Vision as a Behavioural Cue for VR Collaboration}},
    year = {2022},
    journal = {Taiepei 2022: Conference on Computer Supported Cooperative Work and Social Computing, November 12-16, 2022, Taiepei, Taiwan},
    author = {Bovo, Riccardo and Giunchi, Daniele and Muna, Alebri and Steed, Anthony and Costanza, Enrico and Heinis, Thomas},
    number = {1},
    volume = {1},
    publisher = {Association for Computing Machinery},
    doi = {10.1145/3555615}
}

@article{Vrzakova2019,
    title = {{Dynamics of Visual Aention in Multiparty Collaborative Problem Solving using Multidimensional Recurrence antification Analysis}},
    year = {2019},
    journal = {Conference on Human Factors in Computing Systems - Proceedings},
    author = {Vrzakova, Hana and Amon, Mary Jean and Stewart, Angela E.B. and D’Mello, Sidney K.},
    pages = {14},
    publisher = {ACM},
    isbn = {9781450359702},
    doi = {10.1145/3290605.3300572}
}

@inproceedings{Jing2021,
    title = {{EyemR-Vis: Using Bi-Directional Gaze Behavioural Cues to Improve Mixed Reality Remote Collaboration}},
    year = {2021},
    booktitle = {Conference on Human Factors in Computing Systems - Proceedings},
    author = {Jing, Allison and May, Kieran William and Naeem, Mahnoor and Lee, Gun and Billinghurst, Mark},
    month = {5},
    publisher = {Association for Computing Machinery},
    url = {https://doi.org/10.1145/3411763.3451844},
    isbn = {9781450380959},
    doi = {10.1145/3411763.3451844}
}

@article{DAngelo2017,
    title = {{Improving communication between pair programmers using shared gaze awareness}},
    year = {2017},
    journal = {Conference on Human Factors in Computing Systems - Proceedings},
    author = {D'Angelo, Sarah and Begel, Andrew},
    pages = {6245--6255},
    volume = {2017-Janua},
    isbn = {9781450346559},
    doi = {10.1145/3025453.3025573}
}

@article{Zhang2017,
    title = {{Look together: using gaze for assisting co-located collaborative search}},
    year = {2017},
    journal = {Personal and Ubiquitous Computing},
    author = {Zhang, Yanxia and Pfeuffer, Ken and Chong, Ming Ki and Alexander, Jason and Bulling, Andreas and Gellersen, Hans},
    number = {1},
    pages = {173--186},
    volume = {21},
    publisher = {Springer London},
    doi = {10.1007/s00779-016-0969-x},
    issn = {16174909}
}

@inproceedings{Villamor2018,
    title = {{Predicting Successful Collaboration in a Pair Programming Eye Tracking Experiment}},
    year = {2018},
    booktitle = {UMAP 2018 - Adjunct Publication of the 26th Conference on User Modeling, Adaptation and Personalization},
    author = {Villamor, Maureen and Rodrigo, Ma Mercedes},
    number = {July},
    pages = {263--268},
    isbn = {9781450357845},
    doi = {10.1145/3213586.3225234}
}

@article{Schneider2013,
    title = {{Real-Time Mutual Gaze Perception Enhances Collaborative Learning and Collaboration Quality}},
    year = {2013},
    journal = {International Journal of Computer-Supported Collaborative Learning},
    author = {Schneider, Bertrand and Pea, Roy},
    number = {4},
    pages = {375--397},
    volume = {8},
    isbn = {1141201391814},
    doi = {10.1007/s11412-013-9181-4},
    issn = {15561607}
}

@article{Piumsomboon2019,
    title = {{The Effects of Sharing Awareness Cues in Collaborative Mixed Reality}},
    year = {2019},
    journal = {Frontiers Robotics AI},
    author = {Piumsomboon, Thammathip and Dey, Arindam and Ens, Barrett and Lee, Gun and Billinghurst, Mark},
    number = {FEB},
    volume = {6},
    publisher = {Frontiers Media S.A.},
    doi = {10.3389/frobt.2019.00005},
    issn = {22969144}
}

@article{Hong20233D-LLM:Models,
    title = {{3D-LLM: Injecting the 3D World into Large Language Models}},
    year = {2023},
    author = {Hong, Yining and Zhen, Haoyu and Chen, Peihao and Zheng, Shuhong and Du, Yilun and Chen, Zhenfang and Gan, Chuang},
    month = {7},
    url = {http://arxiv.org/abs/2307.12981},
    arxivId = {2307.12981}
}

@inproceedings{Zhang2023ConceptEVA:Summaries,
    title = {{ConceptEVA: Concept-Based Interactive Exploration and Customization of Document Summaries}},
    year = {2023},
    booktitle = {Conference on Human Factors in Computing Systems - Proceedings},
    author = {Zhang, Xiaoyu and Li, Jianping and Chi, Po Wei and Chandrasegaran, Senthil and Ma, Kwan Liu},
    month = {4},
    publisher = {Association for Computing Machinery},
    isbn = {9781450394215},
    doi = {10.1145/3544548.3581260}
}

@inproceedings{Mayer2020EnhancingWorldGaze,
    title = {{Enhancing Mobile Voice Assistants with WorldGaze}},
    year = {2020},
    booktitle = {Conference on Human Factors in Computing Systems - Proceedings},
    author = {Mayer, Sven and Laput, Gierad and Harrison, Chris},
    month = {4},
    publisher = {Association for Computing Machinery},
    isbn = {9781450367080},
    doi = {10.1145/3313831.3376479}
}

@inproceedings{Hindmarsh1998FragmentedEnvironments,
author = {Hindmarsh, Jon and Fraser, Mike and Heath, Christian and Benford, Steve and Greenhalgh, Chris},
title = {{Fragmented Interaction: Establishing Mutual Orientation in Virtual Environments}},
year = {1998},
isbn = {1581130090},
publisher = {Association for Computing Machinery},
address = {New York, NY, USA},
url = {https://doi.org/10.1145/289444.289496},
doi = {10.1145/289444.289496},
booktitle = {Proceedings of the 1998 ACM Conference on Computer Supported Cooperative Work},
pages = {217–226},
numpages = {10},
location = {Seattle, Washington, USA},
series = {CSCW '98}
}

@inproceedings{GuoGRAVL-BERT:Resolution,
    title = {{{GRAVL}-{BERT}: Graphical Visual-Linguistic Representations for Multimodal Coreference Resolution}},
    author = "Guo, Danfeng  and
      Gupta, Arpit  and
      Agarwal, Sanchit  and
      Kao, Jiun-Yu  and
      Gao, Shuyang  and
      Biswas, Arijit  and
      Lin, Chien-Wei  and
      Chung, Tagyoung  and
      Bansal, Mohit",
    booktitle = "Proceedings of the 29th International Conference on Computational Linguistics",
    month = oct,
    year = "2022",
    address = "Gyeongju, Republic of Korea",
    publisher = "International Committee on Computational Linguistics",
    url = "https://aclanthology.org/2022.coling-1.22",
    pages = "285--297",
}

@article{Touvron2023LLaMA:Models,
    title = {{LLaMA: Open and Efficient Foundation Language Models}},
    year = {2023},
    author = {Touvron, Hugo and Lavril, Thibaut and Izacard, Gautier and Martinet, Xavier and Lachaux, Marie-Anne and Lacroix, Timothée and Rozi{\`{e}}re, Baptiste and Goyal, Naman and Hambro, Eric and Azhar, Faisal and Rodriguez, Aurelien and Joulin, Armand and Grave, Edouard and Lample, Guillaume},
    month = {2},
    url = {http://arxiv.org/abs/2302.13971},
    arxivId = {2302.13971}
}

@article{Kottur2021SIMMCConversations,
    title = {{SIMMC 2.0: A Task-oriented Dialog Dataset for Immersive Multimodal Conversations}},
    year = {2021},
    author = {Kottur, Satwik and Moon, Seungwhan and Geramifard, Alborz and Damavandi, Babak},
    month = {4},
    url = {http://arxiv.org/abs/2104.08667},
    arxivId = {2104.08667}
}

@inproceedings{Mahadevan2023Tesseract:Miniature,
    title = {{Tesseract: Querying Spatial Design Recordings by Manipulating Worlds in Miniature}},
    year = {2023},
    booktitle = {Conference on Human Factors in Computing Systems - Proceedings},
    author = {Mahadevan, Karthik and Zhou, Qian and Fitzmaurice, George and Grossman, Tovi and Anderson, Fraser},
    month = {4},
    publisher = {Association for Computing Machinery},
    isbn = {9781450394215},
    doi = {10.1145/3544548.3580876}
}

@inproceedings{Mayer2018TheEnvironments,
author = {Mayer, Sven and Schwind, Valentin and Schweigert, Robin and Henze, Niels},
title = {{The Effect of Offset Correction and Cursor on Mid-Air Pointing in Real and Virtual Environments}},
year = {2018},
isbn = {9781450356206},
publisher = {Association for Computing Machinery},
address = {New York, NY, USA},
url = {https://doi.org/10.1145/3173574.3174227},
doi = {10.1145/3173574.3174227},
booktitle = {Proceedings of the 2018 CHI Conference on Human Factors in Computing Systems},
pages = {1–13},
numpages = {13},
location = {Montreal QC, Canada},
series = {CHI '18}
}

@article{Penedo2023TheOnly,
    title = {{The RefinedWeb Dataset for Falcon LLM: Outperforming Curated Corpora with Web Data, and Web Data Only}},
    year = {2023},
    author = {Penedo, Guilherme and Malartic, Quentin and Hesslow, Daniel and Cojocaru, Ruxandra and Cappelli, Alessandro and Alobeidli, Hamza and Pannier, Baptiste and Almazrouei, Ebtesam and Launay, Julien},
    month = {6},
    url = {http://arxiv.org/abs/2306.01116},
    arxivId = {2306.01116}
}

@article{Yu2022VD-PCR:Resolution,
    title = {{VD-PCR: Improving Visual Dialog with Pronoun Coreference Resolution}},
    year = {2022},
    journal = {Pattern Recognition},
    author = {Yu, Xintong and Zhang, Hongming and Hong, Ruixin and Song, Yangqiu and Zhang, Changshui},
    month = {5},
    volume = {125},
    publisher = {Elsevier Ltd},
    doi = {10.1016/j.patcog.2022.108540},
    issn = {00313203},
    arxivId = {2205.14693}
}

@inproceedings{Sousa2019,
    title = {{Warping deixis: Distorting Gestures to Enhance Collaboration}},
    year = {2019},
    booktitle = {Conference on Human Factors in Computing Systems - Proceedings},
    author = {Sousa, Maurício and Dos Anjos, Rafael Kuffner and Mendes, Daniel and Billinghurst, Mark and Jorge, Joaquim},
    month = {5},
    pages = {1--12},
    volume = {12},
    publisher = {Association for Computing Machinery},
    url = {https://dl.acm.org/doi/10.1145/3290605.3300838},
    address = {New York, NY, USA},
    isbn = {9781450359702},
    doi = {10.1145/3290605.3300838}
}

@inproceedings{Wong2010,
    title = {{Where Are You Pointing? The Accuracy of Deictic Pointing in CVEs}},
    year = {2010},
    booktitle = {Conference on Human Factors in Computing Systems - Proceedings},
    author = {Wong, Nelson and Gutwin, Carl},
    pages = {1029--1038},
    volume = {2},
    publisher = {ACM Press},
    url = {http://portal.acm.org/citation.cfm?doid=1753326.1753480},
    address = {New York, New York, USA},
    isbn = {9781605589299},
    doi = {10.1145/1753326.1753480}
}

@techreport{KongWhatCoreference,
    title = {{What are you talking about? Text-to-Image Coreference}},
    author = {Kong, Chen and Lin, Dahua and Bansal, Mohit and Urtasun, Raquel and Fidler, Sanja}
}

@article{Yu2019WhatDialogues,
    title = {{What You See is What You Get: Visual Pronoun Coreference Resolution in Dialogues}},
    year = {2019},
    author = {Yu, Xintong and Zhang, Hongming and Song, Yangqiu and Song, Yan and Zhang, Changshui},
    month = {9},
    url = {http://arxiv.org/abs/1909.00421},
    arxivId = {1909.00421}
}

@InProceedings{Goel2022WhoNarrations,
    author    = {Goel, Arushi and Fernando, Basura and Keller, Frank and Bilen, Hakan},
    title     = {{Who Are You Referring To? Coreference Resolution In Image Narrations}},
    booktitle = {Proceedings of the IEEE/CVF International Conference on Computer Vision (ICCV)},
    month     = {October},
    year      = {2023},
    pages     = {15247-15258}
}

@article{liu2020evaluating,
  title={{Evaluating the Impact of Virtual Reality on Design Review Meetings}},
  author={Liu, Yifan and Castronovo, Fadi and Messner, John and Leicht, Robert},
  journal={Journal of Computing in Civil Engineering},
  volume={34},
  number={1},
  pages={04019045},
  year={2020},
  publisher={American Society of Civil Engineers}
}

@inproceedings{lopez2018virtual,
  title={{Virtual Reality as a Tool for Emotional Evaluation of Architectural Environments}},
  author={L{\'o}pez-Tarruella Maldonado, Juan and Higuera Trujillo, Juan Luis and I{\~n}arra Abad, Susana and Carmen Llinares Mill{\'a}n, M {\textordfeminine} and Guixeres Provinciales, Jaime and Raya, Mariano Alca{\~n}iz},
  booktitle={Architectural Draughtsmanship: From Analog to Digital Narratives 16},
  pages={889--903},
  year={2018},
  organization={Springer}
}

@incollection{liu2014virtual,
  title={{Virtual Reality to Support the Integrated Design Process: A Retrofit Case Study}},
  author={Liu, Yifan and Lather, Jennifer and Messner, John},
  booktitle={Computing in civil and building engineering (2014)},
  pages={801--808},
  year={2014}
}

@misc{workshopxr,
  title = {{WorkshopXR}},
  author = {{Autodesk}},
  year = {2023},
  howpublished = {\url{https://workshopxr.autodesk.com/}},
  note = {Accessed: [2023-12-09]}
}

@misc{thewild,
  title = {{The Wild}},
  author = {{Autodesk}},
  year = {2023},
  howpublished = {\url{https://thewild.com/}},
  note = {Accessed: [2023-12-09]}
}

@misc{arkio,
  title = {{Arkio}},
  author = {{Arkio ehf.}},  
  year = {2023},
  howpublished = {\url{https://www.arkio.is/}},
  note = {Accessed: [2023-12-09]}  
}

@misc{realwear,
  title = {{RealWear}},
  author = {{RealWear, Inc.}},
  year = {2023},
  howpublished = {\url{https://www.realwear.com/}},
  note = {Accessed: [2023-12-09]}  
}

@article{Kim2020,
author = {Kim, Seungwon and Lee, Gun and Billinghurst, Mark and Huang, Weidong},
doi = {10.1007/s12193-020-00335-x},
issn = {17838738},
journal = {Journal on Multimodal User Interfaces},
month = {jul},
pages = {1--15},
publisher = {Springer},
title = {{The Combination of Visual Communication Cues in Mixed
Reality Remote Collaboration}},
url = {https://doi.org/10.1007/s12193-020-00335-x},
year = {2020}
}

@misc{vrchat,
  title = {{VRChat}},
  author = {{VRChat Inc.}},
  year = {2023},
  howpublished = {\url{https://vrchat.com/}},
  note = {Accessed: [2023-12-09]}
}

@article{wallace2019nlp,
  title={Do NLP models know numbers? probing numeracy in embeddings},
  author={Wallace, Eric and Wang, Yizhong and Li, Sujian and Singh, Sameer and Gardner, Matt},
  journal={arXiv preprint arXiv:1909.07940},
  year={2019}
}

@article{lu2022survey,
  title={A survey of deep learning for mathematical reasoning},
  author={Lu, Pan and Qiu, Liang and Yu, Wenhao and Welleck, Sean and Chang, Kai-Wei},
  journal={arXiv preprint arXiv:2212.10535},
  year={2022}
}

@inproceedings{penzkofer2021conan,
  title={Conan: A usable tool for multimodal conversation analysis},
  author={Penzkofer, Anna and M{\"u}ller, Philipp and B{\"u}hler, Felix and Mayer, Sven and Bulling, Andreas},
  booktitle={Proceedings of the 2021 International Conference on Multimodal Interaction},
  pages={341--351},
  year={2021}
}

@article{lee2024gazepointar,
  title={GazePointAR: A Context-Aware Multimodal Voice Assistant for Pronoun Disambiguation in Wearable Augmented Reality},
  author={Lee, Jaewook and Wang, Jun and Brown, Elizabeth and Chu, Liam and Rodriguez, Sebastian S and Froehlich, Jon E},
  year={2024}
}
